\documentclass[journal=jpclcd,manuscript=letter,layout=traditional]{achemso}
\usepackage{graphicx}
\usepackage{tabularx}
\usepackage{dcolumn}
\usepackage{longtable}
\usepackage{tensor}
\usepackage{placeins}
\usepackage{bm}
\usepackage{amsmath}
\usepackage{amsfonts}
\usepackage{amssymb}
\usepackage{float}
\usepackage{xcolor}%
\providecommand{\U}[1]{\protect\rule{.1in}{.1in}}

\graphicspath{{"C:/Users/GROUP LCPT/Documents/Group/Tomislav/ThirdOrder_FiniteTemperature/figures/"}
{./figures/}{C:/Users/Jiri/Dropbox/Papers/Chemistry_papers/2021/ThirdOrder_FiniteTemperature/figures/}}

\title{Finite-temperature, Anharmonicity, and Duschinsky Effects on the Two-dimensional Electronic Spectra from Ab Initio Thermo-field Gaussian Wavepacket Dynamics}
\author{Tomislav Begu\v{s}i\'{c}}
\email{tomislav.begusic@epfl.ch}
\author{Ji\v{r}\'i Van\'i\v{c}ek}
\email{jiri.vanicek@epfl.ch}
\affiliation{Laboratory of Theoretical Physical Chemistry, Institut des Sciences et
Ing\'enierie Chimiques, Ecole Polytechnique F\'ed\'erale de Lausanne (EPFL),
CH-1015, Lausanne, Switzerland}
\date{\today}

\begin{tocentry}
\includegraphics[scale=1]{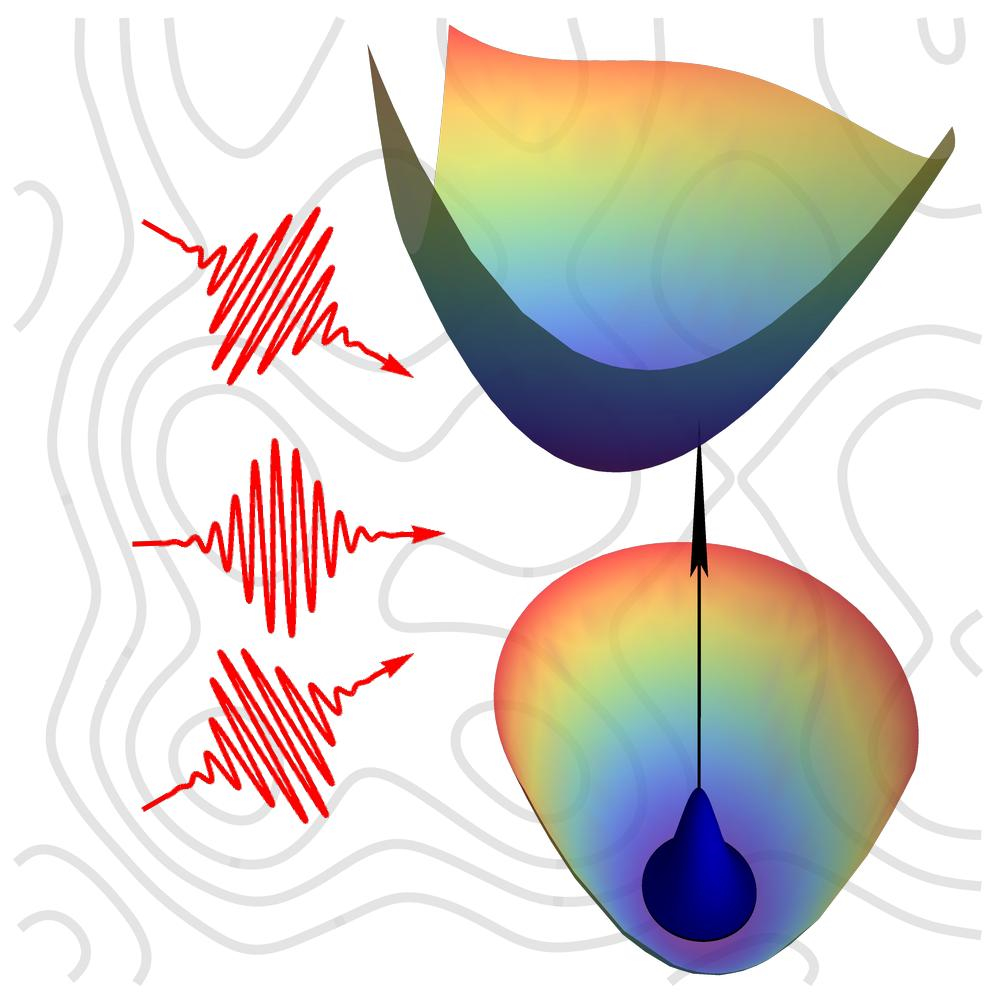}
\end{tocentry}

\begin{document}

\begin{abstract}
Accurate description of finite-temperature vibrational dynamics is
indispensable in the computation of two-dimensional electronic spectra. Such
simulations are often based on the density matrix evolution, statistical
averaging of initial vibrational states, or approximate classical or
semiclassical limits. While many practical approaches exist, they are often of
limited accuracy and difficult to interpret. Here, we use the concept
of thermo-field dynamics to derive an exact finite-temperature expression
that lends itself to an intuitive wavepacket-based interpretation.
Furthermore, an efficient method for computing finite-temperature
two-dimensional spectra is obtained by combining the exact thermo-field
dynamics approach with the thawed Gaussian approximation for the wavepacket
dynamics, which is exact for any displaced, distorted, and Duschinsky-rotated
harmonic potential but also accounts partially for anharmonicity effects in
general potentials. Using this new method, we directly relate a symmetry breaking of the two-dimensional signal to
the deviation from the conventional Brownian oscillator picture.
\end{abstract}

Multidimensional optical spectroscopy is an emerging experimental method for
studying molecular photochemistry and photophysics, but its further
development and the interpretation of new experiments rely heavily on
theoretical modeling.\cite{Yuen-Zhou_Aspuru-Guzik:2011,Kreisbeck_Aspuru-Guzik:2013,book_Zhou_Aspuru-Guzik:2014,Zhou_Zhao:2016,Ikeda_Tanimura:2017,Xiang_Xiong:2018,Conti_Garavelli:2020} To this end, a number of theoretical methods\cite{Kano_Kobayashi:2002,Kobayashi_Otsubo:2007,Fidler_Engel:2013,Bizimana_Turner:2017,Anda_Hansen:2018,Zuehlsdorff_Isborn:2019,
Zuehlsdorff_Isborn:2020,Shedge_Isborn:2021} were developed to account for typical
vibrational-electronic effects occurring in molecular systems, such as
anharmonicity, different curvatures of the ground- and excited-state potential
energy surfaces, or mode-mode mixing (Duschinsky rotation).\cite{Fuji_Kobayashi:2000,Bizimana_Turner:2015,GalestianPour_Hauer:2017,Zhu_Weng:2020,Fumero_Scopigno:2020} In its original formulation, the
second-order cumulant expansion\cite{Mukamel:1982a,book_Mukamel:1999,Segarra-Marti_Rivalta:2018,Picchiotti_Garavelli:2019}
is exact only for the Brownian oscillator
(i.e., displaced harmonic) model and cannot treat the intermode coupling in
the excited state. Although this basic molecular model shaped our
understanding of steady-state, ultrafast, and multidimensional electronic
spectroscopy in the past decades, it is inadequate for many molecules that
exhibit Duschinsky and anharmonicity effects.\cite{Anda_Hansen:2018,Zuehlsdorff_Isborn:2020}
Similar limitations are met when using the semiclassical phase
averaging,\cite{Mukamel:1982, book_Mukamel:1999} also known as the Wigner-averaged classical
limit\cite{Egorov_Rabani:1998,Egorov_Rabani:1999,Shi_Geva:2004,McRobbie_Geva:2009,McRobbie_Geva:2009a} or dephasing
representation.\cite{Mollica_Vanicek:2011,Wehrle_Vanicek:2011,Sulc_Vanicek:2012,Zambrano_Vanicek:2013}
The recently developed third-order cumulant approach seems to overcome these
limitations,\cite{Fidler_Engel:2013, Zuehlsdorff_Isborn:2019,
Zuehlsdorff_Isborn:2020} yet it is accurate only in systems with weakly
coupled or distorted modes.\cite{Zuehlsdorff_Isborn:2020}

In contrast, quantum dynamics
methods\cite{Schubert_Engel:2011,Krcmar_Domcke:2013,Krcmar_Domcke:2015,Picconi_Burghardt:2019,Picconi_Burghardt:2019a,
Begusic_Vanicek:2020a} are well suited for describing the evolution of nuclear
wavepackets but often neglect temperature effects. To avoid the impractical
Boltzmann averaging over the initial states, a number of alternative
strategies for including temperature in wavepacket-based methods have been
proposed.\cite{Matzkies_Manthe:1999,Manthe_Larranaga:2001,Gelman_Kosloff:2003,Nest_Kosloff:2007,Lorenz_Saalfrank:2014,Wang_Zhao:2017,Werther_Grossmann:2018}
We turn to the so-called thermo-field
dynamics,\cite{Suzuki:1985,Takahashi_Umezawa:1996} which transforms the von
Neumann evolution of a density matrix to a Schr\"{o}dinger equation with a
doubled number of degrees of freedom. This approach has only recently been
introduced in chemistry for solving the electronic
structure,\cite{Harsha_Scuseria:2019, Harsha_Scuseria:2019a,
Shushkov_Miller:2019}
vibronic,\cite{Borrelli_Gelin:2016,Borrelli_Gelin:2017,Gelin_Borrelli:2017,Chen_Zhao:2017}
and spectroscopic\cite{Begusic_Vanicek:2020} problems at finite temperature.
Here, we show how it could be used to compute two-dimensional vibronic
spectra. The finite-temperature treatment is combined with the thawed Gaussian
approximation,\cite{Heller:1975} an efficient
first-principles\cite{Wehrle_Vanicek:2014,Wehrle_Vanicek:2015} method for
wavepacket propagation, and applied to the stimulated emission and
ground-state bleach signals of azulene.

In two-dimensional spectroscopy, a nonlinear time-dependent
polarization\cite{Schlau-Cohen_Fleming:2011, book_Mukamel:1999}
\begin{equation}
P^{(3)}(t,t_{2},t_{1})=\left(  \frac{i}{\hbar}\right)  ^{3}\int_{0}^{\infty
}dt^{\prime\prime\prime}\int_{0}^{\infty}dt^{\prime\prime}\int_{0}^{\infty
}dt^{\prime}R^{(3)}(t^{\prime\prime\prime},t^{\prime\prime},t^{\prime}%
)E_{t_1, t_2}(t-t^{\prime\prime\prime})E_{t_1, t_2}(t-t^{\prime\prime}-t^{\prime\prime\prime
})E_{t_1, t_2}(t-t^{\prime}-t^{\prime\prime}-t^{\prime\prime\prime}) \label{eq:P_t}%
\end{equation}
is induced in the sample through interaction with the electric field $E_{t_1, t_2}(t)$
comprised of three light pulses centered at times $-t_{2}-t_{1}$, $-t_{2}$,
and $0$, where $t_{1}$ is the delay between the first two pulses,
$t_{2}$ is the delay between the second and third pulses, and $R^{(3)}(t^{\prime
\prime\prime},t^{\prime\prime},t^{\prime})$ is the third-order response
function.\cite{book_Mukamel:1999} In a heterodyne detection scheme, the
measured signal is\cite{Schlau-Cohen_Fleming:2011}
\begin{equation}
S(t_{3},t_{2},t_{1}) \propto i\int_{-\infty}^{\infty}E_{\text{LO}}(t)P^{(3)}%
(t,t_{2},t_{1})dt,
\end{equation}
where $E_{\text{LO}}(t)$ is the fourth, local oscillator pulse centered at
time $t_{3}$ after the third pulse. The two-dimensional spectrum is obtained
by scanning $S(t_{3},t_{2},t_{1})$ as a function of the three time delays and
Fourier transforming over $t_{1}$ and $t_{3}$. We focus on the absorptive two-dimensional spectrum\cite{Khalil_Tokmakoff:2003,Johnson_Hamm:2017} 
\begin{equation}
\tilde{S}(\omega_3, t_2, \omega_1) = \tilde{S}_{\text{R}}(\omega_3, t_2, \omega_1) +  \tilde{S}_{\text{NR}}(\omega_3, t_2, \omega_1), \label{eq:2D_spec_abs}
\end{equation}
and assume ultrashort and nonoverlapping pulse approximations, where the rephasing and nonrephasing spectra,\cite{Schlau-Cohen_Fleming:2011}
\begin{align}
\tilde{S}_{\text{R}}(\omega_3, t_2, \omega_1) &= \text{Re} \int_0^{\infty} dt_{3}e^{i\omega_{3}t_{3}}\int_{0}^{\infty}dt_{1}e^{-i\omega_{1}t_{1}}%
S_{\text{R}}(t_{3},t_{2},t_{1}), \\
\tilde{S}_{\text{NR}}(\omega_3, t_2, \omega_1) &= \text{Re} \int_0^{\infty} dt_{3}e^{i\omega_{3}t_{3}}\int_{0}^{\infty}dt_{1}e^{i\omega_{1}t_{1}}%
S_{\text{NR}}(t_{3},t_{2},t_{1}),
\end{align}
are defined through
\begin{align}
S_{\text{R}}(t_{3},t_{2},t_{1})  & = C_{1}(t_{1}+t_{2},t_{3},t_{2}+t_{3}) + C_{1}(t_{1},t_{2}+t_{3},t_{3}) - C_{3}(t_{2},t_{3},t_{1}+t_{2}+t_{3})^{\ast}, \label{eq:S_R}\\
S_{\text{NR}}(t_{3},t_{2},t_{1})  & = C_{1}(t_{2},t_{3},t_{1}+t_{2}+t_{3}) + C_{1}(-t_{3},-t_{2}, t_{1}) - C_{3}(t_{1}+t_{2},t_{3},t_{2}+t_{3})^{\ast},\label{eq:S_NR}
\end{align}
and
\begin{equation}
C_i(\tau_{a},\tau_{b},\tau_{c})=\text{Tr}[\hat{\rho}\hat{\mu}_{12}e^{i\hat{H}_{2}%
\tau_{a}/\hbar}\hat{\mu}_{2i}e^{i\hat{H}_{i}\tau_{b}/\hbar}\hat{\mu}_{i2}e^{-i\hat
{H}_{2}\tau_{c}/\hbar}\hat{\mu}_{21}e^{-i\hat{H}_{1}(\tau_{a}+\tau_{b}-\tau
_{c})/\hbar}], \quad i = 1, 3. \label{eq:C_tau}
\end{equation}
In Eq.~(\ref{eq:C_tau}), $\hat{H}_{j}$ are the vibrational Hamiltonians
corresponding to the ground ($j=1$) and excited ($j=2,3$) electronic states,
$\hat{\rho}=\exp(-\beta\hat{H}_{1})/\text{Tr}[\exp(-\beta\hat{H}_{1})]$ is the
vibrational density operator at temperature $T=1/k_{B}\beta$, and $\hat{\mu}_{ij}=\hat{\vec{\mu}}_{ij}\cdot\vec{\epsilon}$ is the electronic transition dipole moment between electronic states $i$ and $j$ projected on the polarization unit vector $\vec{\epsilon}$ of the external electric field. Correlation function $C_1$ corresponds to the stimulated emission and ground-state bleach processes, while $C_3$, which involves a higher excited electronic state, corresponds to the excited-state absorption (see Sec.~1 of the Supporting Information). Although the excited-state absorption term involves, in general, a sum over several higher excited states ($i \geq 3$), here, for the sake of brevity, we consider only one such state.

\begin{figure}
\includegraphics[width=\textwidth]{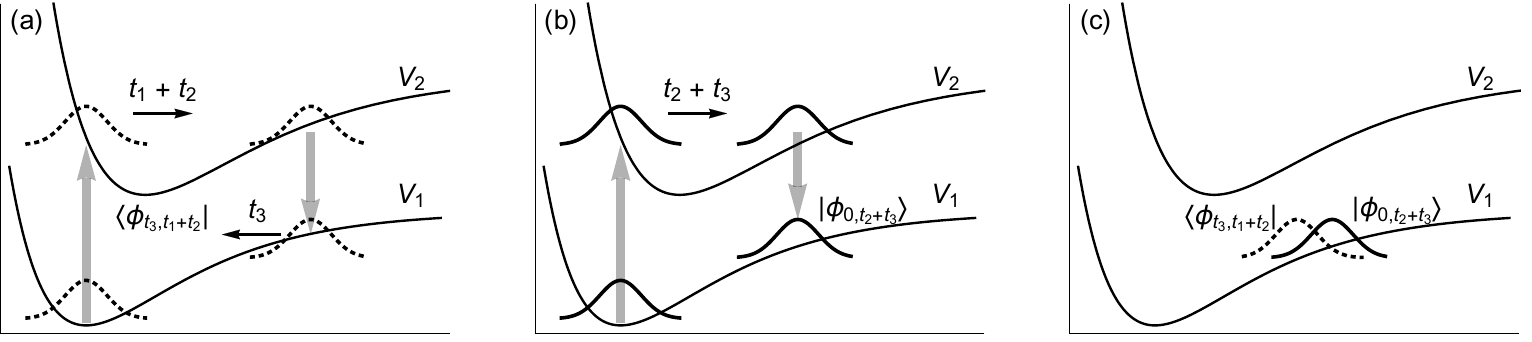} \caption{\label{fig:R_SE_Scheme}Evolution of the bra (a, dotted line) and ket (b, solid line) wavepackets of Eq.~(\ref{eq:C_wp}) for $\tau_a = t_1 + t_2$, $\tau_b = t_3$, and $\tau_c = t_2 + t_3$. Their overlap (c) is stimulated emission term $C_1(t_1 + t_2, t_3, t_2 + t_3)$ [Eq.~(\ref{eq:C_tau})] of rephasing signal $S_{\text{R}}(t_3, t_2, t_1)$ [Eq.~(\ref{eq:S_R})].}
\end{figure}

An intuitive physical interpretation of Eq.~(\ref{eq:C_tau}) is available in
the zero-temperature limit, where the density operator $\hat{\rho}%
=|1,g\rangle\langle1,g|$ is given in terms of the ground vibrational state
$|1,g\rangle$ of the ground electronic state.
Then,\cite{Begusic_Vanicek:2020a}
\begin{equation}
C_i(\tau_{a},\tau_{b},\tau_{c})=\langle\phi_{\tau_{b},\tau_{a}}^{(i)}|\phi_{0,\tau
_{c}}^{(i)}\rangle, \label{eq:C_wp}%
\end{equation}
where
\begin{equation}
|\phi_{\tau,t}^{(i)}\rangle=e^{-i\hat{H}_{i}^{\prime}\tau/\hbar}\hat{\mu}_{i2}%
e^{-i\hat{H}_{2}^{\prime}t/\hbar}\hat{\mu}_{21}|1,g\rangle, \label{eq:phi_t}%
\end{equation}
$\hat{H}_{i}^{\prime}=\hat{H}_{i}-\hbar\omega_{1,g}$, and $\hbar\omega
_{1,g}=\langle1,g|\hat{H}_{1}|1,g\rangle$. In Fig.~\ref{fig:R_SE_Scheme}, we
illustrate how Eq.~(\ref{eq:C_wp}) is evaluated for stimulated emission
contribution $C_1(t_1 + t_2, t_3, t_2 + t_3)$ [Eq.~(\ref{eq:C_tau})] to the rephasing signal [Eq.~(\ref{eq:S_R})]. The bra
nuclear wavepacket is first evolved for a time $\tau_{a}=t_{1}+t_{2}$ in the
excited electronic state and then for a time $\tau_{b}=t_{3}$ in the ground
state; the ket wavepacket is in the ground electronic state during $t_{1}$ and
evolves in the excited state for a time $\tau_{c}=t_{2}+t_{3}$. In general,
during time delays $t_{1}$ and $t_{3}$, also known as coherence and
detection times, the bra and ket wavepackets evolve on different potential
energy surfaces; during the so-called population time $t_{2}$, the two
wavepackets are in the same electronic state: in the ground state for the
ground-state bleach contribution and in the excited electronic state for the
stimulated emission and excited-state absorption components.

We now address the question of whether it is possible to retain
the simple wavepacket picture without neglecting finite-temperature
effects. To answer this question in the
affirmative, we employ thermo-field dynamics, which maps the evolution of a
density operator at finite temperature to the evolution of a wavefunction with
a doubled number of coordinates. In the thermo-field dynamics
theory,\cite{Suzuki:1985} the thermal vacuum is defined as
\begin{equation}
|\bar{0}(\beta)\rangle=\hat{\rho}^{1/2}\sum_{k}|k\tilde{k}\rangle,
\label{eq:0_beta}%
\end{equation}
where $|k\tilde{k}\rangle=|k\rangle|\tilde{k}\rangle$ is the basis vector of
the tensor-product space obtained from the physical (with basis $\{|k\rangle
\}$) and \textquotedblleft fictitious\textquotedblright\ (with basis
$\{|\tilde{k}\rangle\}$) Hilbert spaces. We note that physical operators
(denoted only by hat $\hat{}$, such as $\hat{\rho}$ or $\hat{\mu}$) act only
on the physical subspace. With these definitions, Eq.~(\ref{eq:C_tau}) can be
rewritten as
\begin{equation}
C_i(\tau_{a},\tau_{b},\tau_{c})=\langle\bar{\phi}_{\tau_{b},\tau_{a}}^{(i)}|\bar{\phi
}_{0,\tau_{c}}^{(i)}\rangle, \label{eq:C_tfd}%
\end{equation}
where
\begin{equation}
|\bar{\phi}_{\tau,t}^{(i)}\rangle=e^{-i\hat{\bar{H}}_{i}\tau/\hbar}\hat{\mu
}_{i2}e^{-i\hat{\bar{H}}_{2}t/\hbar}\hat{\mu}_{21}|\bar{0}(\beta)\rangle
\label{eq:phi_bar}%
\end{equation}
is the analogue of $|\phi_{\tau,t}^{(i)}\rangle$ from Eq.~(\ref{eq:phi_t}),
\begin{equation}
\hat{\bar{H}}_{j}=\hat{H}_{j}-\hat{\tilde{H}}_{1} \qquad (j = 1, 2, 3) \label{eq:H_i_bar}%
\end{equation}
is the Hamiltonian acting in the full, tensor-product space, and $\hat
{\tilde{H}}_{1}$ is the ground-state vibrational Hamiltonian acting in the
fictitious space only. The proof of Eq.~(\ref{eq:C_tfd}) goes as follows:
\begin{align}
&C_{i}(\tau_{a},\tau_{b},\tau_{c})   =\langle\bar{0}(\beta)|\hat{\mu}_{12}%
e^{i\hat{\bar{H}}_{2}\tau_{a}/\hbar}\hat{\mu}_{2i}e^{i\hat{\bar{H}}_{i}\tau
_{b}/\hbar}\hat{\mu}_{i2}e^{-i\hat{\bar{H}}_{2}\tau_{c}/\hbar}\hat{\mu}_{21}|\bar
{0}(\beta)\rangle\label{eq:C_tfd_1}\\
&  =\sum_{k_{1},k_{2}}\langle k_{1}\tilde{k}_{1}|\hat{\rho}^{1/2}\hat{\mu
}_{12}e^{i(\hat{H}_{2}-\hat{\tilde{H}}_{1})\tau_{a}/\hbar}\hat{\mu}_{2i}e^{i(\hat{H}%
_{i}-\hat{\tilde{H}}_{1})\tau_{b}/\hbar}\hat{\mu}_{i2}e^{-i(\hat{H}_{2}-\hat
{\tilde{H}}_{1})\tau_{c}/\hbar}\hat{\mu}_{21}\hat{\rho}^{1/2}|k_{2}\tilde{k}%
_{2}\rangle\label{eq:C_tfd_2}\\
&  =\sum_{k_{1},k_{2}}\langle k_{1}|\hat{\rho}^{1/2}\hat{\mu}_{12}e^{i\hat{H}%
_{2}\tau_{a}/\hbar}\hat{\mu}_{2i}e^{i\hat{H}_{i}\tau_{b}/\hbar}\hat{\mu}_{i2}%
e^{-i\hat{H}_{2}\tau_{c}/\hbar}\hat{\mu}_{21}\hat{\rho}^{1/2}|k_{2}\rangle
\langle\tilde{k}_{1}|e^{-i\hat{\tilde{H}}_{1}(\tau_{a}+\tau_{b}-\tau
_{c})/\hbar}|\tilde{k}_{2}\rangle\label{eq:C_tfd_3}\\
&  =\sum_{k_{1},k_{2}}\langle k_{1}|\hat{\rho}^{1/2}\hat{\mu}_{12}e^{i\hat{H}%
_{2}\tau_{a}/\hbar}\hat{\mu}_{2i}e^{i\hat{H}_{i}\tau_{b}/\hbar}\hat{\mu}_{i2}%
e^{-i\hat{H}_{2}\tau_{c}/\hbar}\hat{\mu}_{21}\hat{\rho}^{1/2}|k_{2}\rangle\langle
k_{2}|e^{-i\hat{H}_{1}(\tau_{a}+\tau_{b}-\tau_{c})/\hbar}|k_{1}\rangle
\label{eq:C_tfd_4}\\
&  =\sum_{k_{1}}\langle k_{1}|\hat{\rho}^{1/2}\hat{\mu}_{12}e^{i\hat{H}_{2}\tau
_{a}/\hbar}\hat{\mu}_{2i}e^{i\hat{H}_{i}\tau_{b}/\hbar}\hat{\mu}_{i2}e^{-i\hat{H}%
_{2}\tau_{c}/\hbar}\hat{\mu}_{21}e^{-i\hat{H}_{1}(\tau_{a}+\tau_{b}-\tau_{c}%
)/\hbar}\hat{\rho}^{1/2}|k_{1}\rangle\label{eq:C_tfd_5}\\
&  =\text{Tr}[\hat{\rho}\hat{\mu}_{12}e^{i\hat{H}_{2}\tau_{a}/\hbar}\hat{\mu
}_{2i}e^{i\hat{H}_{i}\tau_{b}/\hbar}\hat{\mu}_{i2}e^{-i\hat{H}_{2}\tau_{c}/\hbar}%
\hat{\mu}_{21}e^{-i\hat{H}_{1}(\tau_{a}+\tau_{b}-\tau_{c})/\hbar}].
\label{eq:C_tfd_6}%
\end{align}
Equation~(\ref{eq:C_tfd_1}) is obtained from Eq. (\ref{eq:C_tfd}) by inserting
the definition~(\ref{eq:phi_bar}) of $\bar{\phi}_{\tau,t}^{(i)}$, while
Eq.~(\ref{eq:C_tfd_2}) results upon substituting relation~(\ref{eq:0_beta}) for
$|\bar{0}(\beta)\rangle$; in going from Eq.~(\ref{eq:C_tfd_2}) to
(\ref{eq:C_tfd_3}) we used the fact that operators acting in different
subspaces commute. In going from (\ref{eq:C_tfd_3}) to (\ref{eq:C_tfd_4}) we
used the conjugation rules relating the physical and
fictitious spaces (see Sec.~2 of the Supporting Information). The resolution of identity and
commutation of $\hat{\rho}^{1/2}$ with $\hat{H}_{1}$ were
used to obtain Eq.~(\ref{eq:C_tfd_5}), and the definition and cyclic property
of the trace to obtain Eq.~(\ref{eq:C_tfd_6}).

Remarkably, the result (\ref{eq:C_tfd}) has exactly the same form as the
zero-temperature expression (\ref{eq:C_wp}) and can be interpreted as in
Fig.~\ref{fig:R_SE_Scheme}. It also allows finite-temperature effects to be
included in regular wavefunction-based codes, by modifying only the definition
of the initial state and the Hamiltonians under which this state is evolved. In Sec.~3 of the Supporting Information, we prove that the same wavepacket picture can be justified even beyond the Born--Oppenheimer approximation, which was invoked implicitly in Eqs.~(\ref{eq:S_R})--(\ref{eq:C_tau}).
To avoid exponentially scaling exact quantum methods on precomputed potential
energy
surfaces\cite{book_MCTDH,Roulet_Vanicek:2019,Choi_Vanicek:2019,Choi_Vanicek:2019a}
or computationally demanding
multiple-trajectory\cite{Tatchen_Pollak:2009,Conte_Ceotto:2020,Buchholz_Ceotto:2016,Buchholz_Ceotto:2017,Curchod_Martinez:2018,Makhov_Shalashilin:2014,Makhov_Shalashilin:2017,Thompson_Martinez:2011,Sulc_Vanicek:2013,Zimmermann_Vanicek:2014,Worth_Burghardt:2004,Richings_Lasorne:2015,Bonfanti_Pollak:2018,Polyak_Knowles:2019,Chen_Zhao:2021}
approaches, we propose using the simple, yet efficient,
single-trajectory thawed Gaussian approximation, which can be interfaced with
on-the-fly ab initio evaluation of potential energy
information.\cite{Vanicek_Begusic:2021}

Let us consider a Gaussian wavepacket
\begin{equation}
\psi_{t}(q) = e^{\frac{i}{\hbar}[(q-q_{t})^{T} \cdot A_{t} \cdot(q-q_{t}) +
p_{t}^{T} \cdot(q-q_{t}) + \gamma_{t}]}, \label{eq:gwp}%
\end{equation}
where $q_{t}$ and $p_{t}$ are the real, $D$-dimensional expectation values of
position and momentum, respectively, $A_{t}$ is a $D \times D$ complex symmetric matrix with
positive-definite imaginary part, $\gamma_{t}$ is a complex scalar whose
imaginary part ensures normalization of the wavepacket, and $D$ is the number
of coordinates. Within the thawed Gaussian approximation,\cite{Heller:1975}
one replaces true potential energy $V(q)$ by its local harmonic
approximation
\begin{equation}
V_{\text{LHA}}(q) = V(q_{t}) + V^{\prime}(q_{t})^{T} \cdot(q-q_{t}) + \frac
{1}{2} (q-q_{t})^{T} \cdot V^{\prime\prime} (q_{t}) \cdot(q-q_{t})
\label{eq:lha}
\end{equation}
about the center $q_{t}$ of the wavepacket, which leads to the following
equations of motion for the Gaussian's
parameters:\cite{Heller:1975,Lasser_Lubich:2020}
\begin{align}
\dot{q}_{t}  &  =m^{-1}\cdot p_{t},\label{eq:q_t_dot}\\
\dot{p}_{t}  &  =-V^{\prime}(q_{t}),\label{eq:p_t_dot}\\
\dot{A}_{t}  &  =-A_{t}\cdot m^{-1}\cdot A_{t}-V^{\prime\prime}(q_{t}%
),\label{eq:A_t_dot}\\
\dot{\gamma}_{t}  &  =L_{t}+\frac{i\hbar}{2}\text{Tr}(m^{-1}\cdot A_{t}),
\label{eq:gamma_t_dot}%
\end{align}
where $L_{t} = p_{t}^{T} \cdot(2 m)^{-1} \cdot p_{t} - V(q_{t})$ is the
Lagrangian along the trajectory $(q_{t}, p_{t})$ and $m$ is the symmetric mass
matrix. According to Eqs.~(\ref{eq:q_t_dot})--(\ref{eq:gamma_t_dot}), the
position and momentum of the Gaussian wavepacket evolve classically, while the
matrix $A_{t}$ depends on the Hessians along the classical trajectory. The
described evolution of the Gaussian wavepacket is exact for a harmonic
potential because the local Taylor expansion of Eq.~(\ref{eq:lha}) becomes
exact in this case. For more general, anharmonic potentials, the method is
only approximate, but typically accurate for moderate anharmonicity and short
times, which makes it practical in spectroscopic
applications.\cite{Heller:1975,Rohrdanz_Cina:2006,Wehrle_Vanicek:2014,Wehrle_Vanicek:2015,Begusic_Vanicek:2018a}
Although the thawed Gaussian propagation is not suited for nonadiabatic dynamics, it can treat accurately the effects that arise due to different force
constants of the ground- and excited-state potential surfaces: mode
distortion, i.e., the change in the frequency of a normal mode, and intermode
coupling or Duschinsky rotation. The on-the-fly ab initio thawed Gaussian
approximation, which uses electronic structure calculations to compute
potential energies, gradients, and Hessians only when needed, was recently
validated for the simulation of finite-temperature
linear\cite{Begusic_Vanicek:2020} and zero-temperature two-dimensional
spectra.\cite{Begusic_Vanicek:2020a}

To construct the initial state, we approximate the ground-state potential
energy surface by a harmonic potential and use the corresponding mass-scaled
normal mode coordinates. Then, in the zero-temperature limit, the initial
state $\psi_{0}(q)=\langle q|1,g\rangle$ is a Gaussian (\ref{eq:gwp}) and
$D=F$, where $F$ is the number of vibrational degrees of freedom. In the
thermo-field dynamics formulation, $D=2F$, the initial state $\bar{\psi}%
_{0}(\bar{q})=\langle\bar{q}|\bar{0}(\beta)\rangle$ is also a Gaussian, and
$\bar{q}=(q,\tilde{q})$ is the $2F$-dimensional coordinate
vector.\cite{Begusic_Vanicek:2020} To solve the equations of motion in the
finite-temperature picture, we need the potential energies, gradients, and
Hessians in the extended coordinate space, which can be easily formulated in
terms of the energies, gradients, and Hessians of the two potential energy
surfaces, as shown in Ref.~\citenum{Begusic_Vanicek:2020}. Remarkably, the
thermo-field dynamics under Hamiltonian $\hat{\bar{H}}_{j}$
[Eq.~(\ref{eq:H_i_bar})] requires exactly the same classical trajectory, in
electronic state $j$, as the conventional, zero-temperature thawed
Gaussian propagation with Hamiltonian $\hat{H}_{j}$
.\cite{Begusic_Vanicek:2020} No further ab initio evaluations are needed for
the finite-temperature implementation, meaning that, within the thawed
Gaussian approximation, the temperature effects can be included almost for
free. The only difference in the computational cost is in solving the
equations of motion with $2F$ rather than $F$ coordinates, which is
approximately $2^{3}=8$ times more expensive due to the roughly cubic scaling
of the involved matrix operations, including matrix-matrix multiplication and
matrix inverse. This cost is, however, negligible compared to the cost of
electronic structure calculations.

Formally, the propagation of the wavepacket according to
Eqs.~(\ref{eq:q_t_dot})--(\ref{eq:gamma_t_dot}) requires not only the
potential energies and gradients but also the Hessians at each step of the
dynamics. In this work, we employed the single-Hessian
method,\cite{Begusic_Vanicek:2019} which further approximates $V^{\prime
\prime}(q_{t})\approx V^{\prime\prime}(q_{\text{ref}})$ in
Eq.~(\ref{eq:A_t_dot}), where $q_{\text{ref}}$ is a reference geometry at
which the Hessian of the excited-state potential surface is evaluated once and
reused during the excited-state dynamics. Because the center of the wavepacket
still follows the fully anharmonic classical trajectory, the single-Hessian
version partially includes anharmonicity effects; in several examples studied
in Ref.~\citenum{Begusic_Vanicek:2019}, the accuracy of this method was shown to be similar to that of the thawed Gaussian approximation. Here, we chose
$q_{\text{ref}}$ as the excited-state minimum. The ground-state potential
surface was assumed to be harmonic in all simulations.

To analyze the effects of the excited-state anharmonicity, we compare the
anharmonic calculations, based on the on-the-fly single-Hessian thawed
Gaussian approximation for the excited-state propagation, with the harmonic
model (also called the generalized Brownian oscillator model), where the
excited-state potential surface is approximated by a harmonic potential fitted
to the surface at its minimum (so-called adiabatic harmonic or adiabatic
Hessian scheme). In the mass-scaled normal mode coordinates of the ground
state, the excited-state force constant is a symmetric, non-diagonal matrix,
whose off-diagonal terms reflect intermode couplings, also known as
Duschinsky mixing. To study the effects of the difference between the excited-
and ground-state force constants on linear and two-dimensional spectra, we
construct the displaced harmonic model (also called the Brownian oscillator
model), where the excited-state force constant is approximated by the force
constant in the ground electronic state. This model neglects mode distortion
and Duschinsky effects. The two-dimensional spectra can be computed exactly
with the thawed Gaussian propagation, as described above, for both harmonic
and displaced harmonic oscillator models. Whereas the exact solution to the
displaced harmonic oscillator model was known before in the form of the
second-order cumulant expansion,\cite{book_Mukamel:1999} to the best of our
knowledge, no method has been published for computing exactly the
two-dimensional spectra of the global harmonic (or generalized Brownian
oscillator) model.\cite{Zuehlsdorff_Isborn:2020}

Azulene is a well-known example of a Kasha-violating
molecule,\cite{Beer_LonguetHiggins:1955} as it emits light from the second,
rather than first, excited electronic state. This is due to the interplay of
two factors:\cite{Prlj_Vanicek:2020} (i) weak nonadiabatic coupling between
S$_{1}$ and S$_{2}$ states and (ii) fast ($\approx1\,$ps) nonradiative decay
from S$_{1}$ to S$_{0}$. These properties make azulene one of the key building
blocks in the synthesis of novel optoelectronic materials.\cite{Xin_Gau:2017}
Although nonadiabatic couplings between the ground and first excited states
play an important role in the photoinduced dynamics of
azulene,\cite{Prlj_Vanicek:2020} they do not affect its vibrationally resolved
S$_{1}\leftarrow\ $S$_{0}$ absorption spectrum. Indeed, the linear absorption
spectrum can be well reproduced using adiabatic, Born-Oppenheimer approaches
that neglect nonadiabatic
effects.\cite{Dierksen_Grimme:2004,Niu_Shuai:2010,Prlj_Vanicek:2020} Here, we
also ignore the nonadiabatic effects on the two-dimensional spectra, which we compute only at short
$t_{2}$ delay times. In the results, we focus on the ground-state bleach and stimulated emission contributions to the two-dimensional spectrum [the first two terms on the right-hand sides of Eqs.~(\ref{eq:S_R}) and (\ref{eq:S_NR})]; according to the oscillator strengths of the S$_1 - \text{S}_{0}$ ($0.009$)\cite{Thulstrup_Michl:1974,Gillispie_Lim:1978,Foggi_Salvi:2003} and S$_2 - \text{S}_{1}$ ($\approx 10^{-5}$)\cite{Gillispie_Lim:1978,Foggi_Salvi:2003} transitions, the excited-state absorption is expected to be $\sim 3$ orders of magnitude weaker.

\begin{figure}
\includegraphics{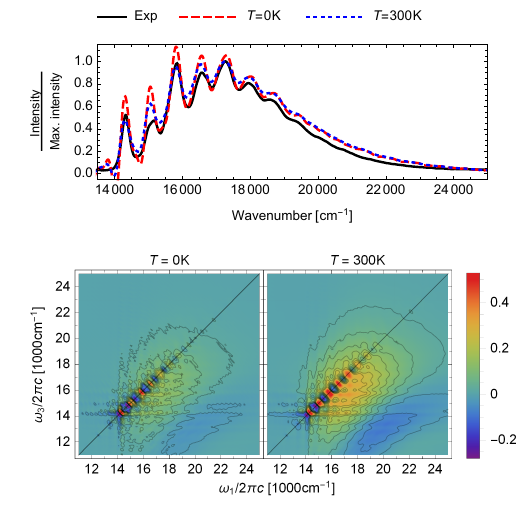} \caption{\label{fig:Spectra_Temperature}Top: S$_1\leftarrow\ $S$_0$ absorption spectra of azulene computed with the on-the-fly ab initio single-Hessian thawed Gaussian approximation at zero temperature (red, dashed) and at $300\,$K (blue, dotted), compared with the experimental spectrum (black, solid) recorded at room temperature in cyclohexane.\cite{Prlj_Vanicek:2020} Bottom: Absorptive two-dimensional electronic spectra [Eq.~(\ref{eq:2D_spec_abs})] at zero delay time ($t_2 = 0$), computed at zero temperature (left) and at $T = 300\,$K (right). Each spectrum shows the sum of the ground-state bleach and stimulated emission terms [first two terms on the right-hand sides of Eqs.~(\ref{eq:S_R}) and (\ref{eq:S_NR})] corresponding to the S$_1$--S$_0$ electronic transition in azulene. See Fig.~S1 for the rephasing and nonrephasing contributions to these spectra and Figs.~S3 and S4 of the Supporting Information for the spectra at delays $t_2 > 0$.}
\end{figure}

In Fig.~\ref{fig:Spectra_Temperature} (top), we compare linear absorption spectra simulated at
$300$ and $0\,$K with the experimental spectrum recorded at room
temperature. One of the main effects of temperature is the broadening of the
spectral features, which also affects the relative intensities of vibronic
peaks, namely, those at $14\,300$ and $15\,800\,$cm$^{-1}$. These
intensities are overestimated in the zero-temperature spectrum but are
corrected by the finite-temperature treatment.

A non-zero temperature has an even stronger effect on the two-dimensional
spectrum (Fig.~\ref{fig:Spectra_Temperature}, bottom). The zero-temperature
spectrum is composed of sharp vibronic peaks, which are broadened and
less resolved in the spectrum computed at $300\,$K. As in the linear spectrum,
the temperature effects modify not only the resolution of the spectrum but
also the relative intensities of the peaks. However, in contrast to the linear
absorption spectrum, where these differences affect only a few peaks and could
still be considered acceptable, the two-dimensional spectrum is strongly
affected due to the increased complexity of spectral features.

\begin{figure}
\includegraphics[scale=1]{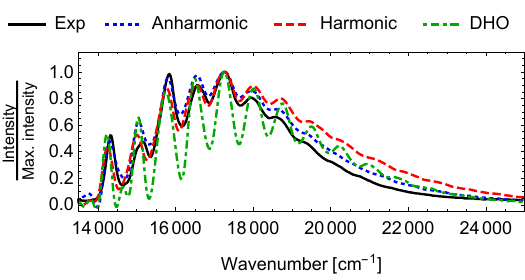} \caption{\label{fig:LinearAbsorption_Anharmonicity}S$_1\leftarrow\ $S$_0$ absorption spectra of azulene computed with the on-the-fly ab initio single-Hessian thawed Gaussian approximation (``Anharmonic'', blue, dotted), harmonic approximation (red, dashed), and the displaced harmonic oscillator (DHO) model (green, dashed-dotted) at $300\,$K, compared with the experimental spectrum (black, solid) recorded at room temperature in cyclohexane.\cite{Prlj_Vanicek:2020}}
\end{figure}

To investigate the effects of anharmonicity, mode distortion, and mode-mode
coupling, we first compare the linear absorption spectra computed using three models
with different accuracies (see Fig.~\ref{fig:LinearAbsorption_Anharmonicity}). The spectrum computed with the
displaced harmonic oscillator model displays a highly regular intensity
pattern, as opposed to the irregular intensities found in the
experiment, and overestimates the frequency spacing between the peaks. The
results are largely improved by including Duschinsky coupling and changes in
the mode frequencies through the global harmonic model. However, the harmonic
approximation suffers from an overly broad tail in the high-frequency region.
This is further corrected by accounting for the anharmonicity effects with the
on-the-fly thawed Gaussian approximation.

\begin{figure}
\includegraphics[scale=0.9]{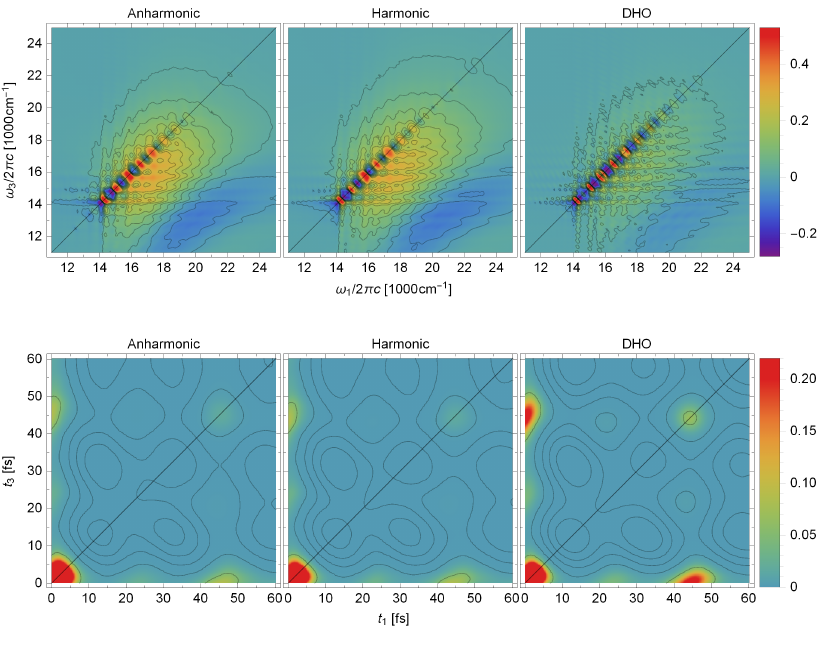} \caption{\label{fig:SpectrumCorr2D_Anharmonicity}Top: Absorptive two-dimensional electronic spectra [Eq.~(\ref{eq:2D_spec_abs})] at zero delay time ($t_2 = 0$), computed with the on-the-fly ab initio single-Hessian thawed Gaussian approximation (``Anharmonic'', left), harmonic approximation (middle), and the displaced harmonic oscillator (DHO) model (right) at $300\,$K. Each spectrum shows the sum of the ground-state bleach and stimulated emission terms [first two terms on the right-hand sides of Eqs.~(\ref{eq:S_R}) and (\ref{eq:S_NR})] corresponding to the S$_1$--S$_0$ electronic transition in azulene. See Fig.~S2 for the rephasing and nonrephasing contributions to these spectra and Figs.~S3 and S4 of the Supporting Information for the spectra at delays $t_2 > 0$. Bottom: First $60\,$fs of $|S_{\text{R}}(t_3, 0, t_1)|$ [Eq.~(\ref{eq:S_R})]. See Fig.~S6 for $|S_{\text{NR}}(t_3, 0, t_1)|$.}
\end{figure}

The corresponding two-dimensional spectra
(Fig.~\ref{fig:SpectrumCorr2D_Anharmonicity}, top) exhibit similar
differences, which we can conveniently analyze in the time domain
(see Fig.~\ref{fig:SpectrumCorr2D_Anharmonicity}, bottom, for $|S_{\text{R}}(t_3, 0, t_1)|$ and Fig.~S6 for $|S_{\text{NR}}(t_3, 0, t_1)|$). The displaced harmonic
oscillator model results in stronger recurrences after $45\,$fs (in $t_{1}$,
$t_{3}$, or in both $t_{1}$ and $t_{3}$) than the harmonic or anharmonic
approaches. This translates into sharper peaks in the two-dimensional
spectrum. The anharmonic spectrum extends less into the high frequency region,
compared to the harmonic and displaced harmonic oscillator models, because the
thawed Gaussian propagation gives a slower initial decay (for $t_{1}, t_{3} <
6\,$fs) in the time domain than the models that neglect anharmonicity (see
Fig.~S5 of the Supporting Information). Subtle differences between the harmonic
and anharmonic excited-state dynamics affect the peak intensities in the
region between $15\,000$ and $18\,000\,$cm$^{-1}$.

Interestingly, for the displaced harmonic oscillator model, $|S_{\text{R}}(t_3, 0, t_1)|$ is symmetric with respect to the diagonal (Fig.~\ref{fig:SpectrumCorr2D_Anharmonicity}, bottom right), which does not hold when mode distortion, rotation, and
anharmonicity are included (Fig.~\ref{fig:SpectrumCorr2D_Anharmonicity},
bottom left and middle). We prove this analytically in the
Supporting Information (Secs.~7 and 8), where we also demonstrate that the asymmetry can appear only in rephasing signal $|S_{\text{R}}(t_3, 0, t_1)|$. Moreover, we show that the (incorrect) symmetry of $|S_{\text{R}}^{\text{DHO}}(t_3, 0, t_1)|$ with respect to the diagonal $t_1 = t_3$ is, more generally, imposed by the second-order cumulant
approximation,\cite{book_Mukamel:1999} which is exact for the displaced
harmonic oscillator model and is employed regularly to model two-dimensional
spectra.\cite{Nenov_Garavelli:2015,Segarra-Marti_Rivalta:2018,Picchiotti_Garavelli:2019,Farfan_Turner:2020}
Hence, the second-order cumulant method cannot account for the asymmetry
induced by the deviation from the displaced harmonic oscillator model. This
erroneous qualitative behavior was difficult to study in the past, partly due to the
absence of practical methods that could easily go beyond the second-order cumulants or Brownian oscillators.

To conclude, we derived a general and exact expression for computing
finite-temperature vibrationally resolved two-dimensional electronic spectra
with wavefunction-based methods. The inclusion of temperature is the key to
simulating spectra of larger systems or solvated molecules, due to the
multitude of low-frequency modes that are thermally excited at room
temperature. By combining the exact expression with the thawed
Gaussian approximation, we developed a practical and efficient method for
computing two-dimensional spectra beyond zero temperature and beyond displaced
harmonic oscillator model. With the help of the newly developed method, we
identified an asymmetry in the time-domain
signal that could serve as evidence for the changes in mode frequencies,
mode-mode coupling, or anharmonicity. This asymmetry cannot be described with
the conventional and widely used second-order cumulant approach.

\section{\label{sec:compdet}Computational Methods}

The ground electronic state of azulene was modeled at the second-order
M{\o }ller--Plesset (MP2) perturbation theory level; the first excited state
was modeled using the second-order Laplace-transformed density-fitted local
algebraic diagrammatic construction [LT-DF-LADC(2)]
scheme,\cite{Kats_Schutz:2009,Ledermuller_Schutz:2013,Ledermuller_Schutz:2014,Schutz:2015}
as implemented in the Molpro 2015 package.\cite{Werner_Wang:2015} cc-pVDZ basis set was used throughout (see Ref.~\citenum{Prlj_Vanicek:2020}). We first
evaluated the Hessians in the ground and excited states at the respective
optimized geometries. Then, starting from the minimum of the ground state, an
on-the-fly ab initio classical trajectory was evolved in the excited
electronic state for 1130 steps with a time step of $8\,$a.u.$\approx0.19\,$fs
(total time $\approx 219\,$fs).

Linear spectra were computed by Fourier transforming the first 500 steps of
the wavepacket autocorrelation function (see
Ref.~\citenum{Vanicek_Begusic:2021}). With regard to the simulation of
two-dimensional spectra, $t_{1}$ and $t_{3}$ times were propagated up to
$\approx106\,$fs (500 steps); $t_{2}$ delays ranged from $0$ (results shown in the main text) to $25\,$fs (130 steps), in intervals of $5\,$fs or 26 steps. Condon approximation, which was justified for the S$_{1}\leftarrow
\ $S$_{0}$ absorption of azulene in Ref.~\citenum{Prlj_Vanicek:2020}, was
employed. Gaussian broadening with a half-width at half-maximum of
$90\,$cm$^{-1}$ was used in both linear and two-dimensional spectra. Linear
spectra were shifted in frequency and scaled in intensity to match at the
maximum intensity peak of the experiment; two-dimensional spectra were shifted
by the same frequency shifts as the linear absorption spectra and scaled according to the maximum of the fully absorptive two-dimensional spectrum [Eq.~(\ref{eq:2D_spec_abs})].

Data supporting this publication can be found at http://doi.org/10.5281/zenodo.4552858.

\begin{acknowledgement}
The authors acknowledge the financial support from the European Research Council (ERC) under the European
Union's Horizon 2020 research and innovation programme (grant agreement No. 683069 -- MOLEQULE) and from the Swiss National Science Foundation through the NCCR MUST (Molecular Ultrafast Science and Technology) Network.
\end{acknowledgement}

\begin{suppinfo}
Derivation of Eqs.~(\ref{eq:S_R})--(\ref{eq:C_tau}), conjugation rules in thermo-field dynamics, thermo-field dynamics expression beyond the Born--Oppenheimer approximation, rephasing and nonrephasing contributions to the spectra of Figs.~\ref{fig:Spectra_Temperature} (bottom) and \ref{fig:SpectrumCorr2D_Anharmonicity} (top), two-dimensional spectra at $t_2 > 0$, $|S_{\text{R}}(t_{3},0,t_{1})|$ at short times, (a)symmetry of the rephasing and nonrephasing spectra in the time domain, and proof of the symmetry of the two-dimensional signal within the second-order cumulant approximation.
\end{suppinfo}

\bibliography{biblio50,additions_FiniteTemperature2D}

\end{document}


\newpage

\graphicspath{{"C:/Users/GROUP LCPT/Documents/Group/Tomislav/ThirdOrder_FiniteTemperature/figures/"}
{./figures/}{C:/Users/Jiri/Dropbox/Papers/Chemistry_papers/2021/ThirdOrder_FiniteTemperature/figures/}}

\section{\label{si_sec:theory_BO}Rephasing and nonrephasing signals for a three-level system within the Born-Oppenheimer approximation}

Since the Eqs.~(6)--(8) of the main text are rarely found in that form, let us summarize how they can be obtained from the expressions that are easily found in the literature. The third-order response function\cite{book_Mukamel:1999, Schlau-Cohen_Fleming:2011}
\begin{equation}
R^{(3)}(t_{3},t_{2},t_{1})=\sum_{\alpha=1}^{4}[R_{\alpha}(t_{3},t_{2},t_{1})-R_{\alpha}(t_{3}%
,t_{2},t_{1})^{\ast}] \label{eq:R}%
\end{equation}
can be expressed in terms of correlation functions
\begin{align}
R_{1}(t_{3},t_{2},t_{1})  &  =C(t_{2},t_{3},t_{1}+t_{2}+t_{3}),\label{eq:R1}\\
R_{2}(t_{3},t_{2},t_{1})  &  =C(t_{1}+t_{2},t_{3},t_{2}+t_{3}),\label{eq:R2}\\
R_{3}(t_{3},t_{2},t_{1})  &  =C(t_{1},t_{2}+t_{3},t_{3}),\label{eq:R3}\\
R_{4}(t_{3},t_{2},t_{1})  &  =C(-t_{3},-t_{2},t_{1}), \label{eq:R4}%
\end{align}
where
\begin{equation}
C(\tau_a, \tau_b, \tau_c) = \text{Tr}[\bm{\hat{\rho}}\bm{\hat{\mu}}e^{i\mathbf{\hat{H}}\tau_a / \hbar}\bm{\hat{\mu}}e^{i\mathbf{\hat{H}}\tau_b / \hbar}\bm{\hat{\mu}}e^{-i\mathbf{\hat{H}}\tau_c / \hbar}\bm{\hat{\mu}}e^{-i\mathbf{\hat{H}}(\tau_a + \tau_b - \tau_c) / \hbar}]. \label{eq:C_tau_gen}
\end{equation}
In Eq.~(\ref{eq:C_tau_gen}), we use the bold font to denote $S \times S$ matrices representing operators acting on the electronic subspace and hat $\hat{\,}$ to denote operators acting on the nuclear subspace; the trace is taken over both nuclear and electronic degrees of freedom. Hence, $\bm{\hat{\rho}} = \exp(- \beta \mathbf{\hat{H}}) /  \text{Tr}[\exp(- \beta \mathbf{\hat{H}})]$ is the full molecular density matrix, $\mathbf{\hat{H}}$ is the molecular Hamiltonian, and $\bm{\hat{\mu}}$ is the dipole moment matrix. Equations~(\ref{eq:R1})--(\ref{eq:C_tau_gen}) can be easily compared to those found in the literature, for example, with Eqs.~(3.5a)--(3.6) of Ref.~\citenum{Schlau-Cohen_Fleming:2011}.

We now consider only three electronic states and invoke the following approximations:
\begin{enumerate}
\item Born-Oppenheimer approximation: $\mathbf{\hat{H}} = \text{diag}(\hat{H}_1, \hat{H}_2, \hat{H}_3)$.
\item Resonance condition: Only transitions between states 1 and 2 or 2 and 3 are allowed:
\begin{equation}
\bm{\hat{\mu}} = \begin{pmatrix}
0 & \hat{\mu}_{21} & 0 \\
\hat{\mu}_{12} & 0 & \hat{\mu}_{23}\\
0 & \hat{\mu}_{32} & 0
\end{pmatrix}.
\end{equation}
\item The system is initially in the ground electronic state, i.e., the temperature effects on the electronic subspace are neglected: $\bm{\hat{\rho}} = \text{diag}(\hat{\rho}, 0, 0)$, where $\hat{\rho} = \exp(-\beta \hat{H}_1) / \text{Tr}[\exp(-\beta \hat{H}_1)]$.
\end{enumerate}
Then, we take the trace over the electronic subspace in Eq.~(\ref{eq:C_tau_gen}) to arrive at
\begin{equation}
C(\tau_a, \tau_b, \tau_c) \approx C_1(\tau_a, \tau_b, \tau_c) + C_3(\tau_a, \tau_b, \tau_c), \label{eq:C_tau_BO}
\end{equation}
where $C_1$ and $C_3$ are defined according to Eq.~(8) of the main text.

Finally, as discussed in Ref.~\citenum{Schlau-Cohen_Fleming:2011}, under the rotating-wave approximation, the rephasing and nonrephasing signals are given by
\begin{align}
S_{\text{R}}(t_3, t_2, t_1) = R_2(t_3, t_2, t_1) + R_3(t_3, t_2, t_1) - R_1(t_3, t_2, t_1)^{\ast}, \label{eq:S_R_ref}\\
S_{\text{NR}}(t_3, t_2, t_1) = R_1(t_3, t_2, t_1) + R_4(t_3, t_2, t_1) - R_2(t_3, t_2, t_1)^{\ast}. \label{eq:S_NR_ref}
\end{align}
Although each $R_{\alpha}$ splits into two terms of the right-hand side of Eq.~(\ref{eq:C_tau_BO}), only one of those terms remains under the rotating-wave approximation. Therefore, Eqs.~(\ref{eq:S_R_ref}) and (\ref{eq:S_NR_ref}) lead to Eqs.~(6) and (7) of the main text.

\section{\label{si_sec:theory_tfd}Conjugation rules in thermo-field dynamics}

In the main text, we have introduced the fictitious Hilbert space with an orthonormal basis $\{|\tilde{k}\rangle\}$, which is related to the physical system (with the orthonormal basis $\{|k\rangle\}$) through the conjugation rules\cite{Suzuki:1985}
\begin{align}
(\hat{A}\hat{B})\, {\tilde{ }} &= \hat{\tilde{A}}\hat{\tilde{B}}, \label{eq:conjg_rule_1a}\\
(\hat{A}| \alpha \rangle) \,{\tilde{ } } &= \hat{\tilde{A}} | \tilde{\alpha} \rangle, \label{eq:conjg_rule_1b}\\
(a \hat{A} + b \hat{B})\,{\tilde{ } } &= a^{\ast} \hat{\tilde{A}} + b^{\ast} \hat{\tilde{B}}, \label{eq:conjg_rule_2} \\
(a | \alpha \rangle + b | \beta \rangle)\,{\tilde{ }} &= a^{\ast} | \tilde{\alpha} \rangle + b^{\ast} | \tilde{\beta} \rangle, \label{eq:conjg_rule_3}
\end{align}
 where the operators ($\hat{A}$, $\hat{B}$) act on and states ($| \alpha \rangle$, $| \beta \rangle$) belong to the original (without tilde) or fictitious (with tilde) Hilbert spaces; $a$ and $b$ are complex numbers.

Let $|\alpha\rangle = \sum_k a_k | k \rangle$. Then, $| \tilde{\alpha} \rangle = \sum_k a_k^{\ast} | \tilde{k} \rangle$,
\begin{equation}
\langle \tilde{\alpha} | \tilde{\beta} \rangle = \sum_k a_k b_k^{\ast} = \langle \beta | \alpha \rangle, \label{eq:conjg_rule_4}
\end{equation}
and
\begin{equation}
\langle \tilde{\alpha}^{\prime} | \hat{\tilde{A}} | \tilde{\alpha} \rangle = \langle \tilde{\alpha}^{\prime} | \hat{\tilde{A}} \tilde{\alpha} \rangle = \langle \hat{A} \alpha |  \alpha^{\prime} \rangle = \langle \alpha | \hat{A}^{\dagger} | \alpha^{\prime} \rangle, \label{eq:conjg_rule_5}
\end{equation}
where in the second step of Eq.~(\ref{eq:conjg_rule_5}) we used Eqs.~(\ref{eq:conjg_rule_1b}) and (\ref{eq:conjg_rule_4}).

We now complete the proof of the main text, namely the step connecting Eqs.~(17) and (18) of the main text, as follows: 
\begin{align}
\langle \tilde{k}_{1}|e^{-i\hat{\tilde{H}}_{1}(\tau_{a}+\tau_{b}-\tau_{c})/\hbar}|\tilde{k}_{2}\rangle 
&=  \langle \tilde{k}_{1}|\left[e^{i\hat{H}_{1}(\tau_{a}+\tau_{b}-\tau_{c})/\hbar}\right]^{\tilde{ }}|\tilde{k}_{2}\rangle \label{eq:tfd_proof_si_1}\\
&=  \langle k_{2}|\left[e^{i\hat{H}_{1}(\tau_{a}+\tau_{b}-\tau_{c})/\hbar}\right]^{\dagger}| k_{1}\rangle  \label{eq:tfd_proof_si_2}\\
&=  \langle k_{2}|e^{-i\hat{H}_{1}(\tau_{a}+\tau_{b}-\tau_{c})/\hbar}| k_{1}\rangle, \label{eq:tfd_proof_si_3}
\end{align}
where in (\ref{eq:tfd_proof_si_1}) we used the conjugation rules (\ref{eq:conjg_rule_1a}) and (\ref{eq:conjg_rule_2}), in going from (\ref{eq:tfd_proof_si_1}) to (\ref{eq:tfd_proof_si_2}) we used Eq.~(\ref{eq:conjg_rule_5}), and in the last step we used the fact that the Hamiltonian is Hermitian.

\section{\label{si_sec:theory_tfd_nonBO}Thermo-field dynamics expression for $C(\tau_a, \tau_b, \tau_c)$ beyond the Born-Oppenheimer approximation}

In the main text, we invoked the Born--Oppenheimer approximation so that the thermo-field dynamics expression could be easily combined with the thawed Gaussian wavepacket dynamics. Of course, in many systems of interest, such as excitonic or multichromophoric systems, or photochemically active molecules, the coupling between electronic states must be included in the calculations. Here, we present the result beyond the Born-Oppenheimer approximation.

As in Eq.~(11) of the main text, we define the thermal vacuum, now in the extended molecular (electronic and vibrational) Hilbert space:
\begin{equation}
|\bar{\mathbf{0}}(\beta) \rangle = \hat{\bm{\rho}}^{1/2} \sum_{k} | \mathbf{k} \tilde{\mathbf{k}} \rangle.
\end{equation}
In other words, the state $|\bar{\mathbf{0}}(\beta) \rangle$ is described not only by a doubled number of vibrational coordinates, but also a doubled number of electronic degrees of freedom. Following similar steps as in the derivation of the main text, we can rewrite Eq.~(\ref{eq:C_tau_gen}) (see Sec. 1 of the Supporting Information) as
\begin{equation}
C(\tau_a, \tau_b, \tau_c) = \langle \bar{\bm{\phi}}_{\tau_b, \tau_a} | \bar{\bm{\phi}}_{0, \tau_c} \rangle, \label{eq:C_tau_gen_TFD}
\end{equation}
where
\begin{equation}
| \bar{\bm{\phi}}_{\tau, t} \rangle = e^{- i \hat{\bar{\mathbf{H}}} \tau / \hbar} \bm{\hat{\mu}} e^{- i \hat{\bar{\mathbf{H}}} t / \hbar} \bm{\hat{\mu}} |\bar{\mathbf{0}}(\beta) \rangle
\end{equation}
and $\hat{\bar{\mathbf{H}}} = \hat{\mathbf{H}} - \hat{\tilde{\mathbf{H}}}$. The result (\ref{eq:C_tau_gen_TFD}) accounts for coupled electronic states and thermal population of excited electronic states. Importantly, Eq.~(\ref{eq:C_tau_gen_TFD}) justifies the thermo-field wavepacket picture even beyond the Born--Oppenheimer approximation.

\section{\label{si_sec:res_R_nonR}Rephasing and nonrephasing two-dimensional spectra}

\begin{figure}[H]
\includegraphics{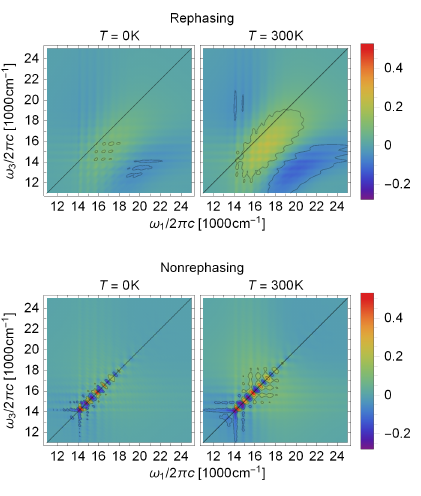} \caption{\label{fig:Spectrum2D_Temperature_R_nonR}Rephasing (top) and nonrephasing (bottom) contributions to the two-dimensional spectra from Fig.~2 of the main text at zero temperature (left) and at $T=300\,$K (right).}
\end{figure}

\begin{figure}[H]
\includegraphics[width=\textwidth]{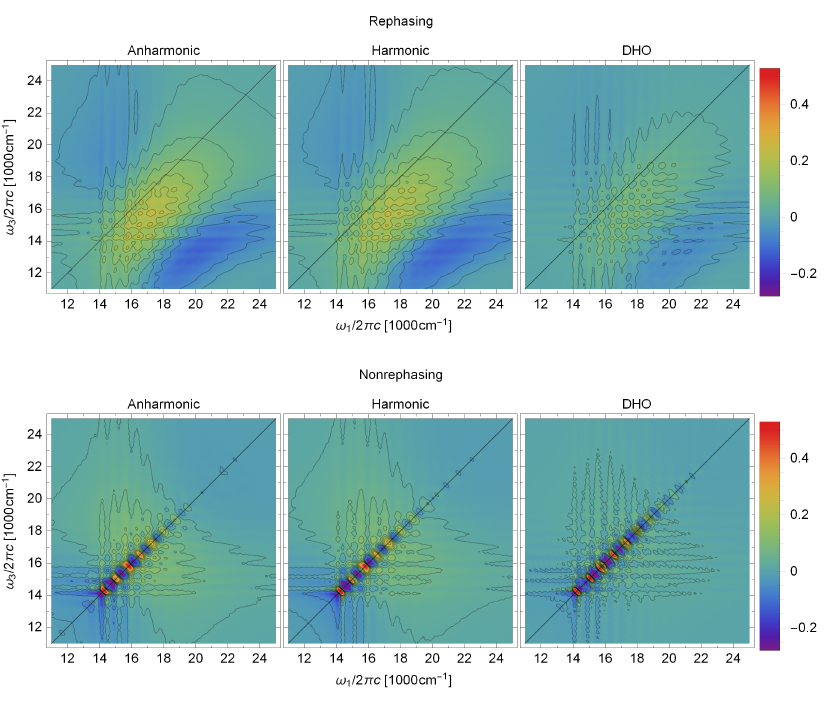} \caption{\label{fig:Spectrum2D_Anharmonicity_R_nonR}Rephasing (top) and nonrephasing (bottom) contributions to the two-dimensional spectra from Fig.~4 of the main text, computed with the on-the-fly ab initio single-Hessian thawed Gaussian approximation (``Anharmonic'', left), harmonic approximation (middle), and the displaced harmonic oscillator (DHO) model (right).}
\end{figure}

\begin{figure}[H]
\includegraphics[width = 1.05\textwidth]{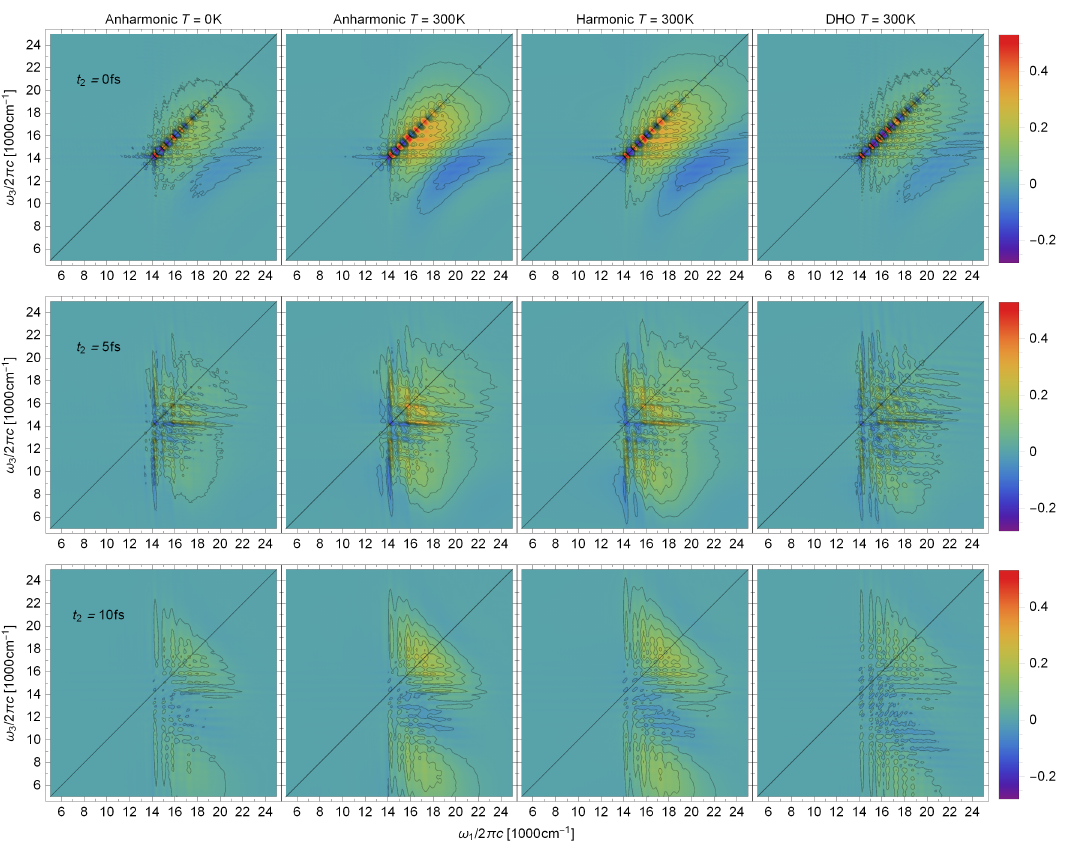} \caption{\label{fig:Spectra_Delays_1}Two-dimensional electronic spectra [Eq.~(3) of the main text] at delay times $t_2 = 0, 5, 10\,$fs, computed with the on-the-fly ab initio single-Hessian thawed Gaussian approximation (``Anharmonic'') at $0\,$K (first column) and $300\,$K (second column), harmonic approximation at $300\,$K (third column), and the displaced harmonic oscillator (DHO) at $300\,$K model (fourth column). Each spectrum shows the sum of the ground-state bleach and stimulated emission terms [first two terms on the right-hand sides of Eqs.~(6) and (7) of the main text] corresponding to the S$_1$--S$_0$ electronic transition in azulene. At zero delay, the ground-state bleach and stimulated emission contributions are indistinguishable. However, already after $5$ or $10\,$fs, the stimulated emission signal moves away from the diagonal, whereas the ground-state bleach spectrum remains close to the diagonal. At later $t_2$ delays (see Fig.~\ref{fig:Spectra_Delays_2}), the stimulated emission starts returning towards the diagonal, reflecting coherent wavepacket dynamics in the excited electronic state. Indeed, the period of about 20--25$\,$fs corresponds to the faster displaced modes (see Table~\ref{tab:azulene_modes}). However, the recurrence is incomplete due to the slower dynamics in the other highly displaced modes.}
\end{figure}

In Figs.~\ref{fig:Spectrum2D_Temperature_R_nonR} and \ref{fig:Spectrum2D_Anharmonicity_R_nonR}, we show separately the rephasing and nonrephasing contributions to the total absorptive spectra of the main text. The rephasing spectrum exhibits a ``checkerboard'' pattern\cite{Borrego-Varillas_Cerullo:2019,Picchiotti_Garavelli:2019}, whereas the nonrephasing peaks appear only along the diagonal.\cite{Begusic_Vanicek:2020a} The interesting nodal structure of the nonrephasing spectrum at $t_2=0$ is a consequence of the broadening. As discussed in Refs.~\citenum{Khalil_Tokmakoff:2003} and \citenum{Schlau-Cohen_Fleming:2011}, the peak lineshape exhibits a ``phase twist'' due to the so-called dispersive component of the broadening function. In the nonrephasing spectrum, the negative intensities about a peak centered at $(\Omega_{1}, \Omega_{3})$ appear at $(\omega_{1} > \Omega_{1}, \omega_{3} > \Omega_{3})$ and $(\omega_{1} < \Omega_{1}, \omega_{3} < \Omega_{3})$ [see, e.g., Fig.~3(b) of Ref.~\citenum{Schlau-Cohen_Fleming:2011}]. By simple addition, these negative features become enhanced in between the diagonal peaks (see Fig.~\ref{fig:Spectrum2D_Anharmonicity_R_nonR}, bottom). Contrary to general knowledge, this nodal structure does not disappear when we sum the rephasing and nonrephasing spectra (see, e.g., Fig.~\ref{fig:Spectra_Delays_1}, top). Due to coherent vibrational dynamics, the nonrephasing and rephasing spectra are different. More precisely, the diagonal peaks of the nonrephasing spectrum are much stronger than the same peaks in the rephasing spectrum and, therefore, their dispersive contributions do not cancel out completely in the total spectrum.\cite{Khalil_Tokmakoff:2003,Schlau-Cohen_Fleming:2011} Consequently, the total spectrum, given by the sum of the rephasing and nonrephasing spectra, is not necessarily composed of purely absorptive peaks. Similar pattern can be seen, for example, in Fig.~4 (leftmost panels showing spectra at $t_2 = 0$) of Ref.~\citenum{Anda_Hansen:2018}.

\section{\label{si_sec:res_t2}Two-dimensional spectra at $t_2 > 0$}

\begin{figure}[H]
\includegraphics[width = 1.05\textwidth]{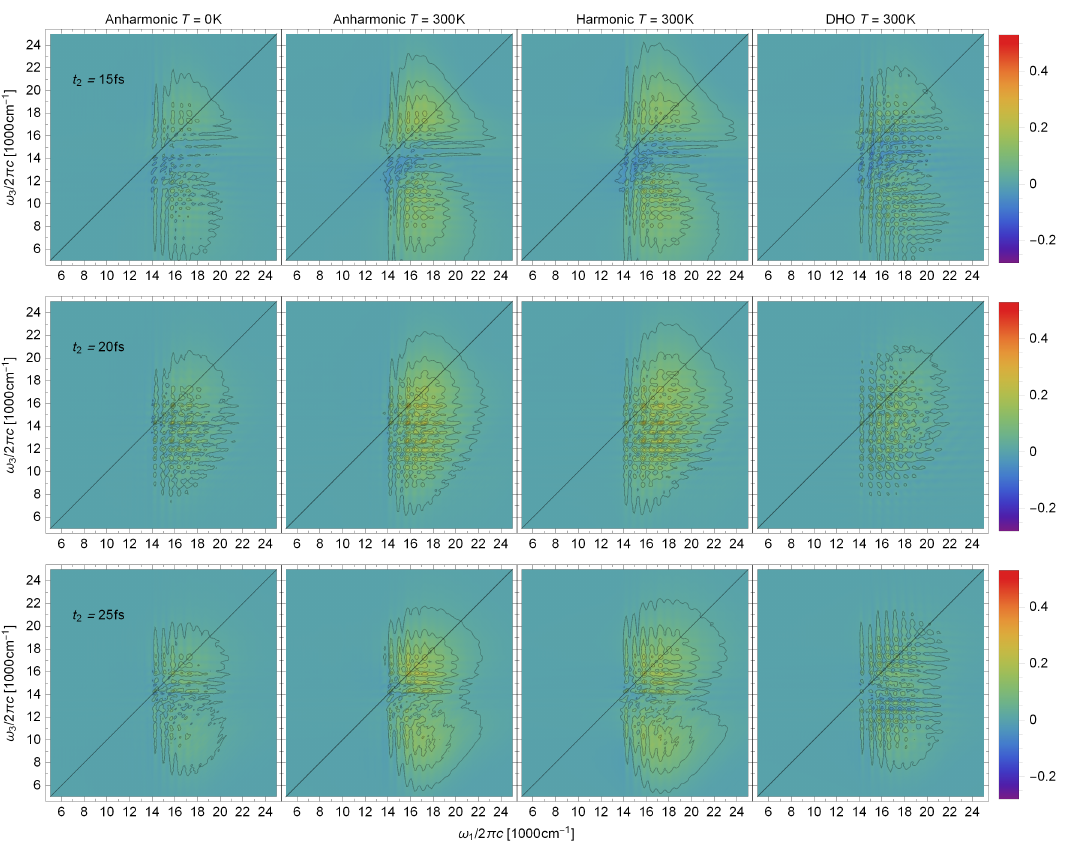} \caption{\label{fig:Spectra_Delays_2}Same as Fig.~\ref{fig:Spectra_Delays_1} but for delay times $t_2 = 15, 20, 25\,$fs.}
\end{figure}

\begin{table}[H]
\caption{\label{tab:azulene_modes}The most displaced modes of azulene. Ground-state frequencies $\omega_i$ of the normal modes are reported in terms of the wavenumber. The dimensionless displacements are defined as $\Delta_i = |\sqrt{\omega_i/\hbar} q_{2,i}|$, where $q_2$ is the excited-state minimum geometry expressed in the mass-scaled normal mode coordinates of the ground electronic state and $q_{2, i}$ is its component along mode $i$. Only the modes with $\Delta_i > 0.25$ are shown.}
\centering
\begin{tabular}{cc}
\hline\hline
Wavenumber / cm$^{-1}$ &  Displacement\\
\hline
 1660 & $ 0.92$ \\
 1490 & $ 0.83$ \\
 1425 & $ 1.22$ \\
 1304 & $ 0.68$ \\
 834  & $ 1.44$ \\
 677  & $ 0.78$ \\
 403  & $ 0.49$ \\
\hline\hline
\end{tabular}
\end{table}

\section{\label{si_sec:res_short_corr}$|S_{\text{R}}(t_3, 0, t_1)|$ at short $t_1$ and $t_3$ times}

\begin{figure}
[H]\includegraphics[width = \textwidth]{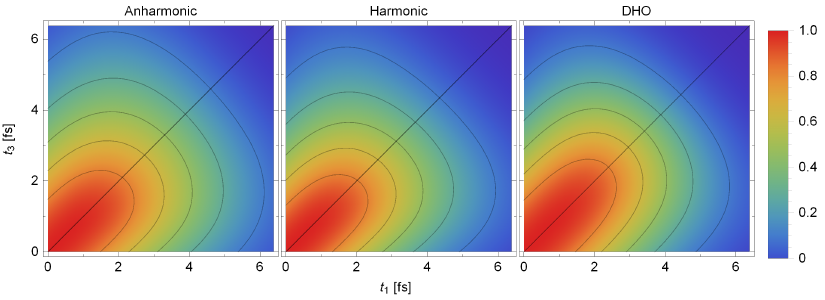} \caption{\label{fig:Correlation2D_Anharmonicity_Short} First $6\ $fs of $|S_{\text{R}}(t_3, 0, t_1)|$ (see Fig.~4 of the main text) computed with the on-the-fly ab initio thawed Gaussian approximation (``Anharmonic''), harmonic approximation, and the displaced harmonic oscillator (DHO) model.}
\end{figure}

\section{\label{si_sec:theory_sym_R_nonR}(A)symmetry of the rephasing and nonrephasing signals in the time domain}

\begin{figure}[H]
\includegraphics[width=\textwidth]{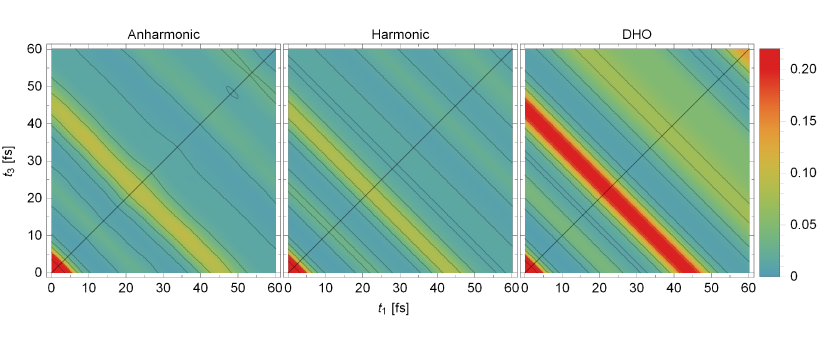} \caption{\label{fig:Correlation2D_Anharmonicity_R_nonR}Absolute value of the nonrephasing time-domain signal ($|S_{\text{NR}}(t_3, 0, t_1)|$) at zero time delay ($t_2 = 0$), including only the ground-state bleach and stimulated emission terms [first two terms on the right-hand side of Eq.~(7) of the main text], computed with the on-the-fly ab initio single-Hessian thawed Gaussian approximation (``Anharmonic'', left), harmonic approximation (middle), and the displaced harmonic oscillator (DHO) model (right) at $300\,$K.}
\end{figure}

In Fig.~\ref{fig:Correlation2D_Anharmonicity_R_nonR}, we show that the absolute value of the nonrephasing signal $S_{\text{NR}}(t_3, 0, t_1)$ at $t_2 = 0$ is symmetric with respect to the diagonal $t_1 = t_3$ for all three---anharmonic, harmonic, and the displaced harmonic oscillator---models, whereas the rephasing signal $|S_{\text{R}}(t_3, 0, t_1)|$ (Fig.~4 of the main text) exhibits this symmetry only for the displaced harmonic oscillator model. Indeed, if we assume the Condon approximation ($\hat{\mu}_{21} \approx \mu = \text{const.}$) and include only the ground-state bleach and stimulated emission contributions, the nonrephasing signal at $t_2 = 0$ can be rewritten as a function of $t_1 + t_3$:
\begin{align}
S_{\text{NR}} (t_3, 0, t_1)
&= C_{1}(0,t_{3},t_{1}+t_{3}) + C_{1}(-t_{3},0, t_{1}) \label{eq:S_NR_sym1}\\
&= 2 |\mu|^4 \text{Tr}[\hat{\rho}e^{i\hat{H}_{1} (t_{1} + t_{3}) /\hbar} e^{-i\hat {H}_{2} (t_1 + t_3) / \hbar}] \label{eq:S_NR_sym2}\\
&= S_{\text{NR}} (t_1, 0, t_3)\label{eq:S_NR_sym3}
\end{align}
and is, therefore, symmetric with respect to the exchange of $t_{1}$ and $t_{3}$. In going from (\ref{eq:S_NR_sym1}) to (\ref{eq:S_NR_sym2}), we used the definition of $C_1$ [Eq.~(8) of the main text], the Condon approximation, the fact that $\hat{\rho}$ commutes with the ground-state vibrational Hamiltonian $\hat{H}_{1}$, and the cyclic property of the trace. The same, however, does not hold for the rephasing contribution:
\begin{align}
S_{\text{R}} (t_3, 0, t_1)
&= 2 C_{1}(t_{1},t_{3},t_{3}) \label{eq:S_R_sym1}\\
&= 2 |\mu|^4 \text{Tr}[\hat{\rho}e^{i\hat{H}_{2} t_{1} /\hbar} e^{i\hat{H}_{1} t_{3}/\hbar} e^{-i\hat {H}_{2} t_{3} /\hbar}e^{-i\hat{H}_{1} t_{1} /\hbar}] \label{eq:S_R_sym2}\\
&\neq 2 |\mu|^4 \text{Tr}[\hat{\rho}e^{i\hat{H}_{2} t_{3} /\hbar} e^{i\hat{H}_{1} t_{1}/\hbar} e^{-i\hat {H}_{2} t_{1} /\hbar}e^{-i\hat{H}_{1} t_{3} /\hbar}] \label{eq:S_R_sym3}\\
&= S_{\text{R}} (t_1, 0, t_3) \label{eq:S_R_sym4}.
\end{align}
Yet, within the displaced harmonic oscillator model, the symmetry is recovered for the absolute value of $S_{\text{R}} (t_1, 0, t_3)$. This is discussed in the following Section.

\section{\label{si_sec:theory_sym_cumulant}Symmetry of the two-dimensional signal within the second-order cumulant approximation}

Within the second-order cumulant method,\cite{book_Mukamel:1999} assuming only two electronic states (i.e., neglecting the excited-state absorption) and the Condon approximation ($\hat{\mu} \approx \mu = \text{const.}$), the components $R_{\alpha}(t_{3},t_{2},t_{1})$ of the third-order response function (\ref{eq:R}) are approximated as
\begin{align}
R_{1}(t_{3},t_{2},t_{1})  &  =|\mu|^4e^{-i\omega_{21}t_{1}-i\omega_{21}t_{3}%
}e^{-g(t_{3})^{\ast}-g(t_{1})-f_{+}(t_{3},t_{2},t_{1})},\label{eq:R_1}\\
R_{2}(t_{3},t_{2},t_{1})  &  =|\mu|^4e^{i\omega_{21}t_{1}-i\omega_{21}t_{3}%
}e^{-g(t_{3})^{\ast}-g(t_{1})^{\ast}+f_{+}(t_{3},t_{2},t_{1})^{\ast}%
},\label{eq:R_2}\\
R_{3}(t_{3},t_{2},t_{1})  &  =|\mu|^4e^{i\omega_{21}t_{1}-i\omega_{21}t_{3}%
}e^{-g(t_{3})-g(t_{1})^{\ast}+f_{-}(t_{3},t_{2},t_{1})^{\ast}}, \label{eq:R_3}%
\\
R_{4}(t_{3},t_{2},t_{1})  &  =|\mu|^4e^{-i\omega_{21}t_{1}-i\omega_{21}t_{3}%
}e^{-g(t_{3})-g(t_{1})-f_{-}(t_{3},t_{2},t_{1})}, \label{eq:R_4}%
\end{align}
where
\begin{equation}
\omega_{21}=\frac{1}{\hbar}\text{Tr}[(\hat{V}_{2}-\hat{V}_{1})\hat{\rho}]
\end{equation}
is the thermally averaged electronic energy gap, $\hat{V}_{i}$ are the
potential energy operators in the ground ($i=1$) or excited ($i=2$) electronic
states,
\begin{equation}
g(t)=\frac{1}{\hbar^{2}}\int_{0}^{t}d\tau_{2}\int_{0}^{\tau_{2}}d\tau
_{1}\text{Tr}[e^{i\hat{H}_{1}\tau_{1}/\hbar}\Delta\hat{V}e^{-i\hat{H}_{1}%
\tau_{1}/\hbar}\Delta\hat{V}\hat{\rho}],
\end{equation}
$\Delta\hat{V}=\hat{V}_{2}-\hat{V}_{1}-\hbar\omega_{21}$, and
\begin{align}
f_{-}(t_{3},t_{2},t_{1})  &  =g(t_{2})-g(t_{2}+t_{3})-g(t_{1}+t_{2}%
)+g(t_{1}+t_{2}+t_{3}),\label{eq:f_minus}\\
f_{+}(t_{3},t_{2},t_{1})  &  =g(t_{2})^{\ast}-g(t_{2}+t_{3})^{\ast}%
-g(t_{1}+t_{2})+g(t_{1}+t_{2}+t_{3}). \label{eq:f_plus}%
\end{align}

For simple systems, such as a pair of harmonic potentials, exact
quantum-mechanical $g(t)$ can be found
analytically.\cite{book_Mukamel:1999,Zuehlsdorff_Isborn:2020} The second-order
cumulant expansion is exact only in the special case of the displaced harmonic
(Brownian) oscillator model. For general potentials, efficient classical
approximations to $g(t)$ can be employed, at the price of introducing
additional approximations in the evaluation of the response function
(\ref{eq:R}). Remarkably, it is easy to show that, no matter how
$g(t)$ is computed, the symmetry relationship
\begin{equation}
|R_{\alpha}(t_{3},t_{2},t_{1})|=|R_{\alpha}(t_{1},t_{2},t_{3})|,\qquad \alpha=1,\ldots
,4,\label{eq:R_i_identity}%
\end{equation}
holds for the four second-order cumulant expressions (\ref{eq:R_1}%
)--(\ref{eq:R_4}). For example, for $R_{3}$ we have
\begin{align}
|R_{3}(t_{3},t_{2},t_{1})|^{2} &  =R_{3}(t_{3},t_{2},t_{1})^{\ast}R_{3}%
(t_{3},t_{2},t_{1})\label{eq:R_3_1}\\
&  =e^{-g(t_{3})^{\ast}-g(t_{1})+f_{-}(t_{3},t_{2},t_{1})}e^{-g(t_{3}%
)-g(t_{1})^{\ast}+f_{-}(t_{3},t_{2},t_{1})^{\ast}}\label{eq:R_3_2}\\
&  =e^{-2\text{Re}[g(t_{3})+g(t_{1})-f_{-}(t_{3},t_{2},t_{1})]}%
\label{eq:R_3_3}\\
&  =e^{-2\text{Re}[g(t_{1})+g(t_{3})-f_{-}(t_{1},t_{2},t_{3})]}%
\label{eq:R_3_4}\\
&  =|R_{3}(t_{1},t_{2},t_{3})|^{2},\label{eq:R_3_5}%
\end{align}
where (\ref{eq:R_3_4}) follows from (\ref{eq:R_3_3}) because
\begin{equation}
f_{-}(t_{3},t_{2},t_{1})=f_{-}(t_{1},t_{2},t_{3}).
\end{equation}
Similar procedure can be followed for other $R_{\alpha}$ functions, where we also
need to use the relation:
\begin{equation}
\text{Re}f_{+}(t_{3},t_{2},t_{1})=\text{Re}f_{+}(t_{1},t_{2},t_{3}).
\end{equation}
Since the exact $R_{\alpha}$ for the displaced harmonic oscillator model can be
written in the form of Eqs.~(\ref{eq:R_1})--(\ref{eq:R_4}), the proof holds in this
special case as well.

Equation~(\ref{eq:R_i_identity}) implies that the symmetry with respect to
exchanging $t_{1}$ and $t_{3}$ is present for arbitrary delay $t_{2}$.
However, in experiments, one cannot measure individual components of the
response function but only their linear combination. For example, the signal measured in the rephasing
phase-matching direction is\cite{Schlau-Cohen_Fleming:2011}
\begin{equation}
S_{\text{R}}(t_{3},t_{2},t_{1})=R_{2}(t_{3},t_{2},t_{1})+R_{3}(t_{3},t_{2},t_{1}).
\end{equation}
Then,
\begin{align}
|S_{\text{R}}(t_{3},t_{2},t_{1})|^{2}  &  =|R_{2}(t_{3},t_{2},t_{1})|^{2}+|R_{3}%
(t_{3},t_{2},t_{1})|^{2}+2\text{Re}[R_{3}(t_{3},t_{2},t_{1})^{\ast}R_{2}%
(t_{3},t_{2},t_{1})]\\
&  \neq |S_{\text{R}}(t_{1},t_{2},t_{3})|^{2}
\end{align}
because
\begin{align}
R_{3}(t_{3},t_{2},t_{1})^{\ast}R_{2}(t_{3},t_{2},t_{1})  &  =e^{-g(t_{3}%
)^{\ast}-g(t_{1})+f_{-}(t_{3},t_{2},t_{1})}e^{-g(t_{3})^{\ast}-g(t_{1})^{\ast
}+f_{+}(t_{3},t_{2},t_{1})^{\ast}}\\
&  =e^{-2g(t_{3})^{\ast}-2\text{Re}[g(t_{1})]+f_{-}(t_{3},t_{2},t_{1}%
)+f_{+}(t_{3},t_{2},t_{1})^{\ast}}\\
&  =e^{-2g(t_{3})^{\ast}-2\text{Re}[g(t_{1})]+2g(t_{2})-2g(t_{2}%
+t_{3})-2\text{Re}[g(t_{1}+t_{2})]+2\text{Re}[g(t_{1}+t_{2}+t_{3})]}\\
&  \neq e^{-2g(t_{1})^{\ast}-2\text{Re}[g(t_{3})]+2g(t_{2})-2g(t_{1}%
+t_{2})-2\text{Re}[g(t_{2}+t_{3})]+2\text{Re}[g(t_{1}+t_{2}+t_{3})]}\\
&  =e^{-2g(t_{1})^{\ast}-2\text{Re}[g(t_{3})]+f_{-}(t_{1},t_{2},t_{3}%
)+f_{+}(t_{1},t_{2},t_{3})^{\ast}}\\
&  = R_{3}(t_{1},t_{2},t_{3})^{\ast}R_{2}(t_{1},t_{2},t_{3}).
\end{align}
Nevertheless, the equality is restored at $t_{2}=0$ because $R_{2}%
(t_{3},0,t_{1})=R_{3}(t_{3},0,t_{1})$:
\begin{equation}
|S_{\text{R}}(t_{3},0,t_{1})|^{2}=4|R_{3}(t_{3},0,t_{1})|^{2}=4|R_{3}(t_{1},0,t_{3}%
)|^{2}=|S_{\text{R}}(t_{1},0,t_{3})|^{2}.
\end{equation}

\bibliography{biblio50,additions_FiniteTemperature2D}